# On the energy spectrum of the one-dimensional Klein-Gordon Oscillator

Nagalakshmi A. Rao

*Department of Physics, Government Science College,*

*Bangalore-560001,Karnataka, India.*

drnarao@gmail.com

B. A. Kagali

*Department of Physics, Jnanabharathi Campus, Bangalore University,*

*Bangalore-560056, India.*

bakagali@hotmail.com

In the present article, we describe a method of introducing the harmonic potential into the Klein-Gordon equation, leading to genuine bound states. The eigenfunctions and eigenenergies are worked out explicitly.

**PACS** 03.65 Ge; 03.65 Pm.

**Keywords** KG Oscillator, eigenenergy, Weber's equation, confluent hypergeometric functions.

# 1   Introduction

The harmonic oscillator is one of the most important quantum systems, as it is one of the very few that can be solved exactly. Extension of the harmonic oscillator problem to relativistic domain is interesting in itself. Motivation comes from various situations, where relativistic effects play an important role. Relativistic quantal oscillators are indeed significant, as special features are seen when the oscillator motion becomes relativistic. While several authors[1-5] have addressed the relativistic oscillator in one dimension, Rao et.al. have analysed the Dirac oscillator in two spatial dimension, extending the prescription of Moshinsky[6]. Recently Mirza et.al.[7] have studied the Klein-Gordon and Dirac oscillators in a non-commutative space. On the other hand, Villalba[8] has discussed the eigenfunctions of the 2+1 Dirac oscillator using polar coordinates. Witwit[9] has presented the energy spectrum of the non-relativistic two dimensional anharmonic oscillator by the Hill determinant method. In the present article, we explore the salient features of the Klein-Gordon oscillator.

# 2   The Klein-Gordon equation with the oscillator potential

The one dimensional Klein-Gordon equation for a vector potential[10] is

$$\left\{\frac{d^2}{dx^2} + \frac{(E-V(x))^2 - m^2c^4}{c^2\hbar^2}\right\}\psi(x) = 0. \tag{1}$$

The above equation may be rewritten in the Schrodinger form, with an effective energy and effective potential as

$$\frac{d^2}{dx^2} + (E_{eff} - V_{eff})\psi = 0 \tag{2}$$

with $E_{eff} = \frac{E^2 - m^2c^4}{c^2\hbar^2}$   and   $V_{eff} = \frac{2EV(x) - V^2(x)}{c^2\hbar^2}$.

When the harmonic oscillator potential $V(x) = \frac{1}{2}mw^2x^2$, with $m$, the mass of the particle and $w$, the classical frequency is introduced in the usual manner, the effective potential takes the form

$$V_{eff} = \frac{Emw^2x^2 - \frac{1}{4}m^2w^4x^4}{c^2\hbar^2}. \tag{3}$$

Lipas[3] has clearly pointed out that there will be no true bound states for positive energy particles since this $V_{eff}$ is unbounded from below at infinity. In view of this ill-behaving nature, we propose a novel approach to describe the Klein-Gordon Oscillator, motivated by the prescription of Moshinsky. Further, we show that the eigenenergies resulting from such a treatment have an appropriate non-relativistic limit and the eigenfunctions resemble those of the non-relativistic oscillator.

## The Klein-Gordon Oscillator

The one-dimensional free particle Klein-Gordon equation may be written as

$$\left( (p_x)_{op}^2 - \frac{E^2 - m^2 c^4}{c^2} \right) \psi = 0 \tag{4}$$

with the operator, $(p_x)_{op} = -i\hbar \frac{\partial}{\partial x}$. Following the prescription of Mirza et.al[8], we introduce an interaction that couples to the momentum as follows

$$p_x \rightarrow p_x - imwx \tag{5}$$

and

$$p_x^\dagger = p_x + imwx, \tag{6}$$

where $p_x^\dagger$ is the adjoint of the momentum operator $p_x$. In the presence of the coupling, we replace the operator $p_x^2$ by the symmetric operator $\frac{1}{2}(p_x p_x^\dagger + p_x^\dagger p_x)$. The Klein-Gordon equation in this prescription reads

$$\left[ \frac{d^2}{dx^2} - \frac{m^2 w^2 x^2}{\hbar^2} + \left\{ \frac{E^2 - m^2 c^4}{c^2 \hbar^2} \right\} \right] \psi = 0. \tag{7}$$

It is readily seen that in the limit of $mc^2 >> \hbar w$ the resulting Klein-Gordon equation reduces to the non-relativistic harmonic oscillator equation and hence Equation(7) may be thought of as the Klein-Gordon Oscillator. In what follows, we set

$$k^2 = \frac{E^2 - m^2 c^4}{c^2 \hbar^2}. \tag{8}$$

With $\lambda = \frac{mw}{\hbar}$, Eqn.(7) becomes

$$\frac{d^2 \psi}{dx^2} + \left( k^2 - \lambda^2 x^2 \right) \psi = 0 \tag{9}$$

which we identify as the Weber's differential equation[11].

The above equation may be transformed into

$$y \frac{d^2 \psi}{dy^2} + \frac{1}{2} \frac{d\psi}{dy} + \left( \frac{k'}{2} - \frac{1}{4} y \right) \psi = 0 \tag{10}$$

with the substitution $y = \lambda x^2$ and

$$k' = \frac{k^2}{2\lambda} = \frac{E^2 - m^2c^4}{2mc^2\hbar w}. \tag{11}$$

In order to rewrite Equation(10) in the standard form, we split off the asymptotic solution, and hence try a solution of the form $\psi(y) = exp(-\frac{y}{2}) \, \phi(y)$, which leads to

$$y\frac{d^2\phi}{dy^2} + \left(\frac{1}{2} - y\right)\frac{d\phi}{dy} + \left(\frac{k'}{2} - \frac{1}{4}\right)\phi = 0. \tag{12}$$

We identify the above equation as the Kummer's differential equation[12]. Interstingly, this equation is structurally similar to that of the Schrodinger equation with the usual harmonic oscillator potential[13].

The eigenfunctions may be expressed in terms of regular confluent hypergeometric functions $M(a, c, y)$ as

$$\phi(y) = AM(a, \frac{1}{2}, \, y) + By^{\frac{1}{2}} \, M(a + \frac{1}{2}, \, \frac{3}{2}, y) \tag{13}$$

where A and B are arbitrary constants and

$$a = -\left(\frac{k'}{2} - \frac{1}{4}\right). \tag{14}$$

In terms of the original variable $x$, the wavefunctions may be written as

$$\psi(x) = Ae^{-\frac{mw}{2\hbar}x^2} \, M(a, \, \frac{1}{2}, \, \lambda x^2) + Be^{-\frac{mw}{2\hbar}x^2}\sqrt{\frac{mw}{\hbar}}x \, M(a + \frac{1}{2}, \, \frac{3}{2}, \, \lambda x^2). \tag{15}$$

Physically admissible solutions require finiteness and normalisability and the necessary square integrability of $\psi$ implies the vanishing of the wave function at infinity. Nonetheless, it is seen from the properties of hypergeometric functions that for large values of $y$, $M(a, c, y)$ is an exponentially growing function. This obviously means that the normalisation integral diverges. However, if the hypergeometric series terminates to a polynomial, normalisation is ensured, leading further to quantisation of energy[13].

## Energy Spectrum

The energy eigenvalues of spin zero particles bound in this oscillator potential may be found using Eqn.(14). Following Greiner, we see that $a = -n$ for even states, and $a + \frac{1}{2} = -n$ for odd states, $n$ being a non-negative integer. We consider the two cases separately and work out the energy eigenvalues explicitly.

**Even states:** The even states alone are present when $a = -n$ and $B = 0$ with $n = 0, 1, 2, 3, ...$ refering to the ground, first, second, third, etc., even states respectively.

Using Equation(14), it is straightforward to see that the even state eigenenergies are given by

$$E^2 = m^2c^4 + 2\left(2n + \frac{1}{2}\right)mc^2\hbar w. \tag{16}$$

It would be interesting to calculate the binding energy of a relativistic oscillator defined as
$$e = E - mc^2 \tag{17}$$
and compute the relativistic correction for different classical frequencies. It is readily seen from Eqn.(16) that in the first order approximation

$$e \approx (2n + \frac{1}{2})\hbar w.$$

In the second order approximation of Eqn.(16), we obtain

$$E^{even} \approx mc^2 + (2n + \frac{1}{2})\hbar w - \frac{1}{2}\left(2n + \frac{1}{2}\right)^2 \frac{\hbar^2 w^2}{mc^2}. \tag{18}$$

It is seen that the total energy is the sum of the rest energy, a non-relativistic term and a small relativistic correction.

**Odd states:** The odd states alone are present when $a + \frac{1}{2} = -n$ and $A = 0$ with $n = 0, 1, 2, 3, ...$ refering to the first, second, third, etc., odd states respectively. As before, using Equation(14), we see that the odd state eigenenergies are given by

$$E^2 = m^2c^4 + 2\left(2n + \frac{3}{2}\right)mc^2\hbar w. \tag{19}$$

While first order approximation of the above equation gives us

$$e \approx (2n + \frac{3}{2})\hbar w,$$

the second order approximation would yield

$$E^{odd} \approx mc^2 + (2n + \frac{3}{2})\hbar w - \frac{1}{2}\left(2n + \frac{3}{2}\right)^2 \frac{\hbar^2 w^2}{mc^2}. \tag{20}$$

Combining Equations(18) and (20) to include both the even and odd states, we obtain

$$E^2 = m^2c^4 + 2\left(n + \frac{1}{2}\right)mc^2\hbar w, \tag{21}$$

further leading to

$$E_n \approx mc^2 + (n + \frac{1}{2})\hbar w - \frac{1}{2}\left(n + \frac{1}{2}\right)^2 \frac{\hbar^2 w^2}{mc^2}. \tag{22}$$

with the index $n = 0, 1, 2, 3, \ldots$. It is trivial to note that in the non-relativistic limit, the binding energy of the Klein-Gordon oscillator is

$$e \approx (n + \frac{1}{2})\hbar w,$$

in agreement with the energy of the non-relativistic harmonic oscillator.

**Eigenfunctions**: As a consequence of the well-behaved convergence and normalisability, termination of the hypergeometric series enables us to express the eigenfunctions in terms of the well-known Hermite polynomials. The confluent hypergeometric functions are related to the Hermite polynomials through the following equations:

$$H_{2n}(\xi) = (-1)^n \frac{(2n)!}{n!} M(-n, \frac{1}{2}, \xi^2) \tag{23}$$

$$H_{2n-1}(\xi) = (-1)^n \frac{2(2n+1)!}{n!} M(-n, \frac{3}{2}, \xi^2) \tag{24}$$

where $\xi = \sqrt{\lambda} x$ is implied. In view of the above equations, the even and the odd eigenfunctions may be expressed as

$$\psi_{even} = N_n e^{(-\lambda/2)x^2} H_{2n}(\sqrt{\lambda} x) \tag{25}$$

$$\psi_{odd} = N_n e^{(-\lambda/2)x^2} H_{2n+1}(\sqrt{\lambda} x) \tag{26}$$

with the normalisation constant given by

$$N_n = \sqrt{\sqrt{\frac{\lambda}{\pi}} \frac{1}{2^n n!}} \tag{27}$$

However, the even and odd eigenfunctions may be combined and the stationary states of the relativistic oscillator are

$$\psi_n(x) = \sqrt{\sqrt{\frac{\lambda}{\pi}} \frac{1}{2^n n!}} exp(-\frac{1}{2}\lambda x^2) H_n(\sqrt{\lambda} x) \tag{28}$$

## 3  Results and Discussion

In the relativistic situation, it is seen that if the oscillator potential is included as a Lorentz vector, the effective potential is unbounded from below at infinity and no true bound states occur. Extending the prescription of Moshinsky, the Klein-Gordon equation leads to an oscillator with genuine bound states. Interestingly, this prescription yeilds the relativistic eigenenergies having unequal spacing. While discreteness is a property of bound states, equispacedness is a characteristic feature of non-relativistic oscillator. At high energies, the oscillator becomes more sluggish and the spacing between the levels change slowly compared to the non-relativistic oscillator. We have computed the eigenenergies of the relativistic oscillator for different strength parameter (b). It is seen from the table of energies that the KG oscillator has an appropriate non-ralativistic limit.

It is worthwhile extending the calculation to the three dimensional case, the results of which may be compared with those of the 3D isotropic Dirac oscillator. Furthermore, we notice that relativistic anharmonic oscillators may be generated modifying our prescription. Nevertheless, owing to the complexity of the potential, analytical solutions may not be possible and numerical techniques are imperative and such studies are worth undertaking.

## Acknowledgements

One of the authors (Nagalakshmi A.Rao) gratefully acknowledges the University Grants Commission and the Department of Collegiate Education in Karnataka for the award of Teacher Fellowship under the Faculty Improvement Programme. Thanks are extended to Ms. M. V. Jayanthi, IAS, Commissioner for Collegiate Education in Karnataka for her endearing encouragement and support.

# Energy eigenvalues of the 1D KG Oscillator

$$\left(b = \frac{\hbar w}{mc^2}\right)$$

|   | b=0.1 | | b=0.001 | | b=0.0001 | |
|---|---|---|---|---|---|---|
| n | $\bar{E}_R$ | $\bar{E}_{Nr}+1$ | $\bar{E}_R$ | $\bar{E}_{Nr}+1$ | $\bar{E}_R$ | $\bar{E}_{Nr}+1$ |
| 0 | 1.09545 | 1.05 | 1.001 | 1.0005 | 1.0001 | 1.00005 |
| 1 | 1.18322 | 1.15 | 1.002 | 1.0015 | 1.0002 | 1.00015 |
| 2 | 1.26491 | 1.25 | 1.003 | 1.0025 | 1.0003 | 1.00025 |
| 3 | 1.34164 | 1.35 | 1.00399 | 1.0035 | 1.0004 | 1.00035 |
| 4 | | | 1.00499 | 1.0045 | 1.0005 | 1.00045 |
| 5 | | | 1.00598 | 1.0055 | 1.0006 | 1.00055 |
| 6 | | | 1.00698 | 1.0065 | 1.0007 | 1.00065 |
| 7 | | | 1.00797 | 1.0075 | 1.0008 | 1.00075 |
| 8 | | | 1.00896 | 1.0085 | 1.0009 | 1.00085 |
| 9 | | | 1.00995 | 1.0095 | 1.0010 | 1.00095 |
| 10 | | | 1.01094 | 1.0105 | 1.0011 | 1.00105 |
| 20 | | | 1.02078 | 1.0205 | 1.0021 | 1.00205 |
| 30 | | | 1.03053 | 1.0305 | 1.0031 | 1.00305 |
| 31 | | | 1.03150 | 1.0315 | 1.00315 | 1.00310 |
| 50 | | | | | 1.00509 | 1.00505 |
| 70 | | | | | 1.00707 | 1.00705 |
| 80 | | | | | 1.00807 | 1.00805 |
| 90 | | | | | 1.00906 | 1.00905 |
| 100 | | | | | 1.01005 | 1.01005 |